\documentstyle[11pt,newpasp,twoside,epsf]{article}

\begin{document}

\title{Theoretical considerations on colliding clumped winds}

\author{Rolf Walder}
\affil{Institute of Astronomy, ETH Z\"urich, Switzerland
       http://www.astro.phys.ethz.ch/staff/walder/walder.html}

\author{Doris Folini}
\affil{Observatoire de Strasbourg, Universit\'{e} Louis Pasteur, 
       Strasbourg, France}


\begin{abstract}
First attempts are made to derive astrophysical implications of the
collision of clumped stellar winds from order of magnitude estimates
and preliminary numerical simulations. Compared to colliding smooth
winds, we find that the most significant differences occur in widely
separated systems like WR~140. Clumped winds de-stabilize the wind-wind
interaction zone of such systems. Highly compressed, cold knots of
WR-wind material can form. Hydrogen rich material is likely to be
mixed into these knots by the excited turbulence. Such knots,
therefore, are good candidates to form dust. We briefly discuss to
what degree our results can be applied to other systems and look at
different possibilities for the origin and nature of the
inhomogeneities in hot star winds.
\end{abstract}


\keywords{hot massive stars, colliding winds, stability, dust}

\section{Introduction}
It is now established that WR-winds are inherently highly structured:
they carry substantial density- and velocity-inhomogeneities (see
e.g. Cherepashchuk 1990; L\'{e}pine, Eversberg, \& Moffat 1999, and
references therein). Recently, traces of such inhomogeneities have
also been found in OB-winds (Eversberg, L\'{e}pine, \& Moffat 1998;
Howk et al. 2000) and LBV-winds (Grosdidier et al. 1998).  The results
of Grosdidier et al. (1998) also indicate that the wind is structured
to large distances. The question arises how the interaction of winds
carrying such inhomogeneties takes place and whether -- and to what
degree -- we have to revise the theory of colliding winds (see
e.g. Folini and Walder 2000 for a review of the present theory). In
this paper, we attempt to make some first steps towards answering
these questions, knowing well, however, that many more have to
follow. Most of our argumentation is still on the level of basic
estimates and only a few quantitative results will be presented. We
mainly concentrate on colliding winds in hot star binaries. Similar
arguments will be valid, however, for other colliding wind binaries
and for wind driven structures in general.

Some ideas on the nature of the inhomogeneities are briefly sketched
in Section~2. In Section~3 we discuss the physics of the interaction
of single wind-inhomogeneities with a stationary wind-wind-interaction
zone, before we generalize that picture in Section~4 to the collision
of structured winds. We finally draw some conclusions in Section~5.

\section{The nature of the inhomogeneities in hot star winds}
\label{sec:nature_inhomo}
Presently, there is no coherent theoretical model of the formation of
density- and velocity-inhomogeneities in hot star winds, of their
number, size, and distribution. The best attempt in this direction so
far has been presented in a series of papers by Owocki and co-workers
(e.g. Owocki, Castor, \& Rybicki 1988, for a review see Feldmeier
\& Owocki 1998). This work has shown that winds from hot massive stars
(and line-driven winds in general) are inherently unstable. Within the
frame of the applied approach -- spherical symmetry and sampled
line-opacities -- small disturbances in one of the flow variables grow
very rapidly. A series of high-density, shock-confined sheets is
formed. The sheets move outwards in the otherwise strongly rarefied
wind. However, these 1D models can neither explain the above referred
observations in the optical and UV showing outmoving clumps, nor can
the associated shocks account for the observed X-ray flux of hot
massive stars.

The results of Owocki and co-workers can be interpreted as a form of
supersonic turbulence, restricted to spherical symmetry. It is a very
general feature of highly compressible turbulence that the flow is
structured into high-density knots and filaments in combination with
large voids. (see e.g. Walder \& Folini 2000 for the density
distribution function of turbulence driven by planar colliding
flows). One can hope that the multidimensional generalization of the
above 1D-results leads to the formation of knots and shocks within the
atmosphere which better fit the observations. The radiative line
forces then not only accelerate the wind but would also, in
combination with compressible effects, force the turbulence to form
the knots. It will be the task of future work to determine the density
distribution function within such a framework.

What role in this framework would possible inhomogeneities at the
wind-base play that are provided by the interior dynamics of the
star? A likely scenario is that they are soon completely erased. The
wind-inhomogeneities of the outer wind would then be mainly the result
of the complex interplay between the stellar radiation field and the
turbulent flow. The 1D results of Owocki and co-workers point in this
direction. On the other hand, there seem to be cases where such
excitations at the wind-base -- perhaps particularly large ones --
leave their fingerprints on the wind throughout the whole atmosphere,
modifying the turbulence persistently. First attempts by Owocki (1998)
to model DACs point in this direction.

Dense clumps in the wind could, in principle, also be formed by
locally strongly enhanced cooling. However, such a scenario is very
unlikely in the presence of the strong stellar radiation
field. Density enhancements of only a factor of ten would require to
cool the wind locally to temperatures as low as 3000~K. Moreover, the
formation of dense clumps by cooling would immediately disturb the
wind, and turbulence would be exciteed also in this case.

\section{Interaction of a single clump with a wind-wind interaction
         zone}
\label{sec:single_clump}
Simulations of colliding clumped winds of entire systems are out of
reach even for todays computer resources. Moreover, as discussed in
the last section, the exact distribution of the clumps is not yet
known. We thus follow another approach and discuss single, different,
important physical effects of clumped wind collisions before trying to
unify the picture by bringing the different effects together. When
being specific, we use parameters typical for WR~140 at periastron
(Williams et al. 1990, 1995)\footnote{Separation $d_{p}
\approx 2.35$~AU, WR-wind: $\dot{M} = 5.7 \cdot
10^{-5}$~M$_{\odot}$/y, $v_{\infty} = 2860$~km/s, O-star-wind:
$\dot{M} = 1.8 \cdot 10^{-6}$~M$_{\odot}$/y, $v_{\infty} =
3200$~km/s. For a smooth, stationary wind, the density of the WR-wind
immediately before it gets shocked is on the order of $10^9$~cm$^{-3}$.
We further assume a wind temperature of 35'000~K, a value which is not
crucial for the following.}. Most of the discussion will hold,
appropriately applied, for narrow systems as well. Some differences
will, however, be discussed in the next section.

\paragraph{Collisions of clumps and voids with a shock in the frame
           of 1D-Euler-equations}

\begin{figure}[tbp]
\centerline{ \plottwo{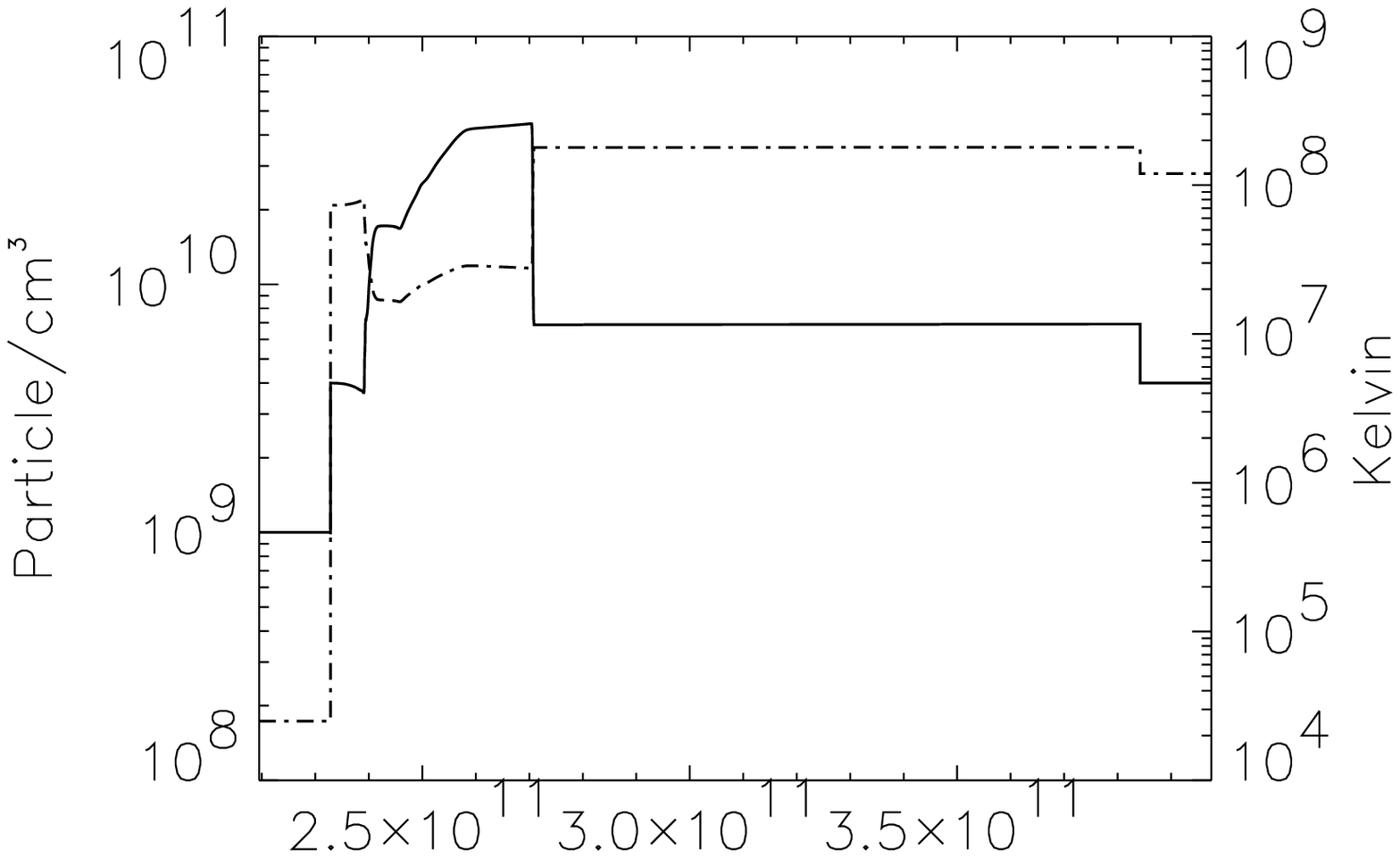}{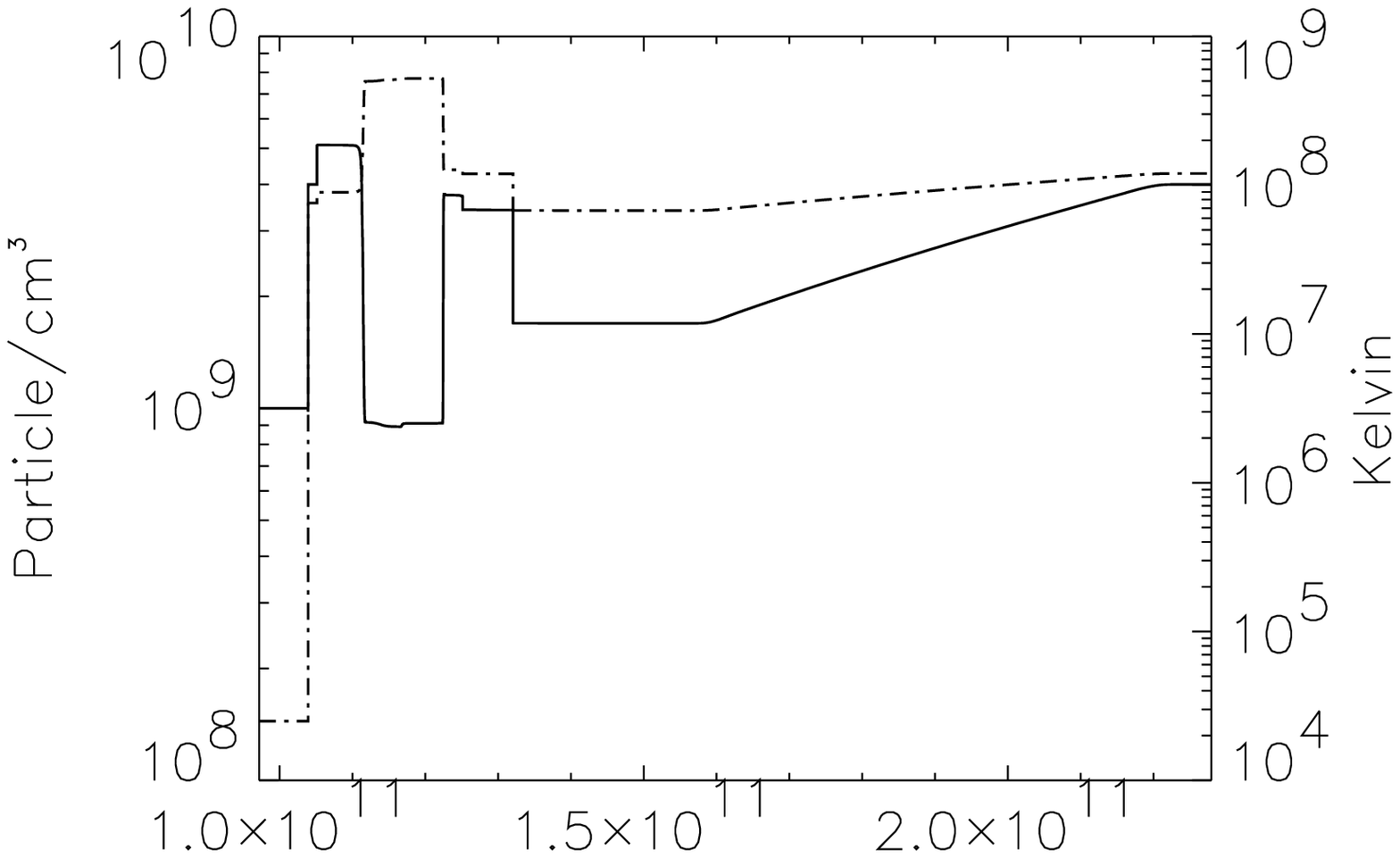} }
\caption{Wind-wind interaction zone after the interaction of a 
         clump (left) and a void (right) with a stationary shock in a
         1D test-simulation. The WR star is to the left, the O-star to
         the right. The original position of the undisturbed shock was
         at $1.31 \cdot 10^{11}$~cm (arbitrary position) and the clump
         or the void respectively was injected on the WR-side of the
         shock. Shown are density (solid) and temperature profiles
         (dashed). The post-shock values of the undisturbed flow are:
         $N = 4 \cdot 10^9$~cm$^{-3}$, $v = 715$~km/s, and $T =
         1.2\cdot 10^8$~K. The clump causes the shock to move in the
         direction of the O-star, the void, on the contrary, towards
         the WR-star. Several waves -- shocks and rarefactions -- are
         created and travel through the interaction zone, leading to a
         complex density- and temperature-structure.}

\label{fig:clump_shock_1d}
\end{figure}

Figure~\ref{fig:clump_shock_1d} shows the interaction of a
clump (with a density-enhancement of a factor 10) and a void (with a
density-deficit of factor 10) with a shock which is stationary with
respect to the undisturbed WR-wind. In both cases, the shock
immediately becomes unstationary and secondary waves start to travel
in the wind-wind interaction zone (WWIZ). In 2D or 3D, these secondary
waves eventually de-stabilize the interface separating the
WR-star-material from the O-star-material when interacting with it.

Another important point is that already a single void or clump leads
temporarily to a complex density- and temperature-structure of the
WWIZ when interacting with it. In particular, as the shock is hit by a
clump and thus is pushed forward, the shock strength decreases,
resulting in a significantly lower temperature and higher density of
the post-shock gas as compared to the stationary value (as low as
$3\cdot 10^7$~K in the example). In the case of voids, on the other
hand, the shock is strengthened, leading to a significantly higher
post-shock temperature. Thus, the inhomogeneous structure of the
WR-wind directly translates into a great variation of cooling times of
the now patchy WWIZ. Denser regions with low temperature can cool in
less than 1000 seconds, tenuous regions with high temperature need
more than 0.1 years.
%
%
%
%
%
%
%
This has to be compared to 2 days, the cooling time for the stationary
value, to 9 hours, the time a sound wave (corresponding to a
temperature of $1.2\cdot 10^8$ ~K) needs to travel through the width
of the WWIZ, and to 2.6 days, the time a sound wave needs to cross
the separation between the two stars.

3D effects will not qualitatively alter the {\it conclusion} from this
1D example: Even in systems as widely separated as WR~140 parts of the
shocked WR-wind can cool in the very center of the system\footnote{At
least in the short time interval near periastron. In apastron, typical
wind-densities are about a factor of 10 lower. With cooling times 100
times longer even dense clumps have no chance to cool in the center
and can thus also not be compressed to high densities.}. This is in
strong contrast to the case of the collision of smooth flows
investigated by Stevens, Blondin \& Pollack (1992). The WWIZ is
unstable and has a patchy structure. Cooled parts are highly
compressed to densities of about $10^{13}$~cm$^{-3}$ in the case of
WR~140.

\paragraph{Heat conduction}
Heat conduction by thermal electrons and, if optically very thick, by
X-ray photon diffusion, is a very efficient process. The heat
conduction coefficient is a very non-linear function of temperature,
$\kappa \propto T^{\beta}$, with $\beta=2.5$ for thermal electrons and
$\beta \approx 3$ for photon diffusion (see e.g. Spitzer 1962). 

{\it Confining shocks:} Only briefly we want to note that heat
conduction lowers significantly the temperature of the WWIZ by heating
the stellar winds upwind of the shocks confining the WWIZ. In
addition, the density of the hot parts of the WWIZ is much higher than
for non-heat-conducting shocks. These two effects are enough
to diminish the cooling time so much that the shocks most probably
become radiative in almost all colliding wind binaries. However, even
small magnetic fields, which are likely to be present, reduce this
effect to potentially zero. An extended discussion can be found in
Myasnikov \& Zhekov (1998).

{\it Clumps embedded in hot material:} 
Heat conduction certainly plays an important role with regard to the
dense clumps floating in the very hot environment of the patchy WWIZ.
Even when a significant magnetic field threads the clumps, there
always exists a direction where the ultrafast electrons of the hot
phase can penetrate the clump, heating it up. However, as long as it
is optically thin, the dense clump can radiate this energy again. A
1D-estimate of the heating-cooling-balance reads
\begin{equation}
\frac{\kappa_0}{3.5} \frac{\partial^2 T^{3.5} }{\partial x^2} =
N^2 \Lambda(T) .
\end{equation}
If we apply $\Lambda = 1.5 \cdot 10^{-19} T^{-0.5}$ [erg cm$^6$/s],
$\kappa_0 = 6 \cdot 10^7 $  [erg cm$^2$/(s K$^{7/2}$)], and make the
approximation
$
\partial^2 T / \partial x^2 
\approx T_{high}^2 / \left( \Delta x \right)^2, 
$
we obtain, as on order of magnitude estimate, for the thickness of a
cooling, heat-conducting front between the hot, shocked gas and the
embedded clump,
\begin{equation}
\Delta x \approx 10^{10} \left(\frac{10^{10}}{N_{front}} \right)
                         \left(\frac{T_{high}}{10^7}\right)^2 
         \hspace{0.5cm} \mbox{ [cm] } .
\end{equation}
Typical values (we consider WR140 again) for $T_{high}$ are between
$10^7$ and $10^8$~K. For $N_{front}$, a mean density in the front,
typical values are between $10^{11}$ and $10^{13}$~cm$^{-3}$. Thus, a
typical thickness of a heat-conducting, cooling front is between
$10^8$ and $10^{10}$~cm. Cool, condensed clumps embedded in hot gas
must be bigger than this scale for not being evaporated by conductive
heating.

{\it We conclude} that big enough clumps -- clumps greater than
several solar radii in the WR-wind -- can also cool when
heat-conduction is considered. Moreover, they will not be evaporated
by heat conduction but act like a `catalyst', enhancing the cooling of
the hot parts of the WWIZ: energy is transported from hot parts to
cold, dense parts where it can be radiated. In addition, if dense
clumps of WR-material float in the shocked O-star wind, hydrogen rich
material is cooled and thus becomes part of the clumps. This material
eventually diffuses into the carbon rich material elsewhere in the
clump. Unfortunately, diffusion is an inefficient process. However,
clumps eventually collide with each other or with a shock. Then, the
clump-interior becomes turbulent, the material is stirred, and the
surface through which hydrogen can diffuse is much bigger. Mixing of
the two materials is likely in this case.

\paragraph{Equilibrium time-scales}

In a pure, completely ionized hydrogen gas, the electron
self-collision time (the time electrons need to thermalize after a
significant disturbance like a shock-passage) can be estimated by 
(Spitzer 1962)
\begin{equation}
t_c \left(e-e\right) \approx 0.2664 \times \frac{T^{3/2}}{N_e \ln \Lambda} [s].
\end{equation}
For the self-collision time $t_c\left(p-p\right)$ of the protons and
the equilibration time $t_c\left(p-e\right)$ between electrons and
protons the relations
\begin{equation}
t_c\left(e-e\right) : 
t_c\left(p-p\right) : 
t_c\left(p-e\right) \approx
1 : \left(\frac{m_p}{m_e}\right)^{1/2} : \frac{m_p}{m_e} = 
1 : 43 : 1836
\end{equation}
hold. For typical values of the shocked gas of WR~140 in periastron,
$N_e = 10^{10}$~cm$^{-3}$, $T = 10^8$~K, one obtains $t_c
\left(e-e\right) \approx 1$~s, and $t_c \left(e-p\right)
\approx 2000$~s. For a pure, completely ionized helium gas, the
equipartition time is of the same order of magnitude. The
equipartition time is thus comparable with the estimates for the
fastest cooling time-scale and even of the same order of magnitude as
sound crossing times of the high temperature regions of the WWIZ.

{\it We conclude} that for quantitative predictions two-temperature
models are required, i.e. models which distinguish between the electron
temperature, $T_e$, and the ion temperature, $T_{ion}$. 
We add two notes: 1) In a patchy interaction zone, electron
temperatures higher than $5 \cdot 10^8$~K (see Figure~1) are likely in
some parts of the WWIZ, at least if they have time to
thermalize. Then, a considerable amount of relativistic electrons is
present. Such electrons may contribute to non-thermal emission in the
WWIZ. 2) In situations where the ion-electron equipartition time is
comparable to typical transport times, shock-preheating by heat
conducting thermal electrons will be significantly suppressed.
\section{Interaction of inhomogeneous winds}
\label{sec:supersonic_turb}
From the estimates presented above and from test examples it becomes
clear that a quantitatively correct, consistent model of colliding
inhomogeneous winds is out of reach even with present day computer
resources. In this section we will try to develop a qualitative
picture of those effects which are important in an astrophysical
sense. Moreover, we try to draw some observational consequences.

In a multidimensional context, the size of clumps will be small
compared to the shock surface. Thus, not the entire shock will be
pushed forward when colliding with a clump, but only a little section
of it. At the same time, parts of the shock are colliding with clumps
of different sizes, but also with voids. Moreover, the same part of
the shock will collide subsequently with clumps and with voids. In a
certain moment of time, the shape of the WWIZ thus looks very chaotic
and its patchy interior is governed by turbulence. On longer terms and
in a statistical sense, however, the dynamics of the WWIZ of colliding
clumped flows will not necessarily differ much from the case of
colliding smooth flows.
\begin{figure}[t]
\plottwo{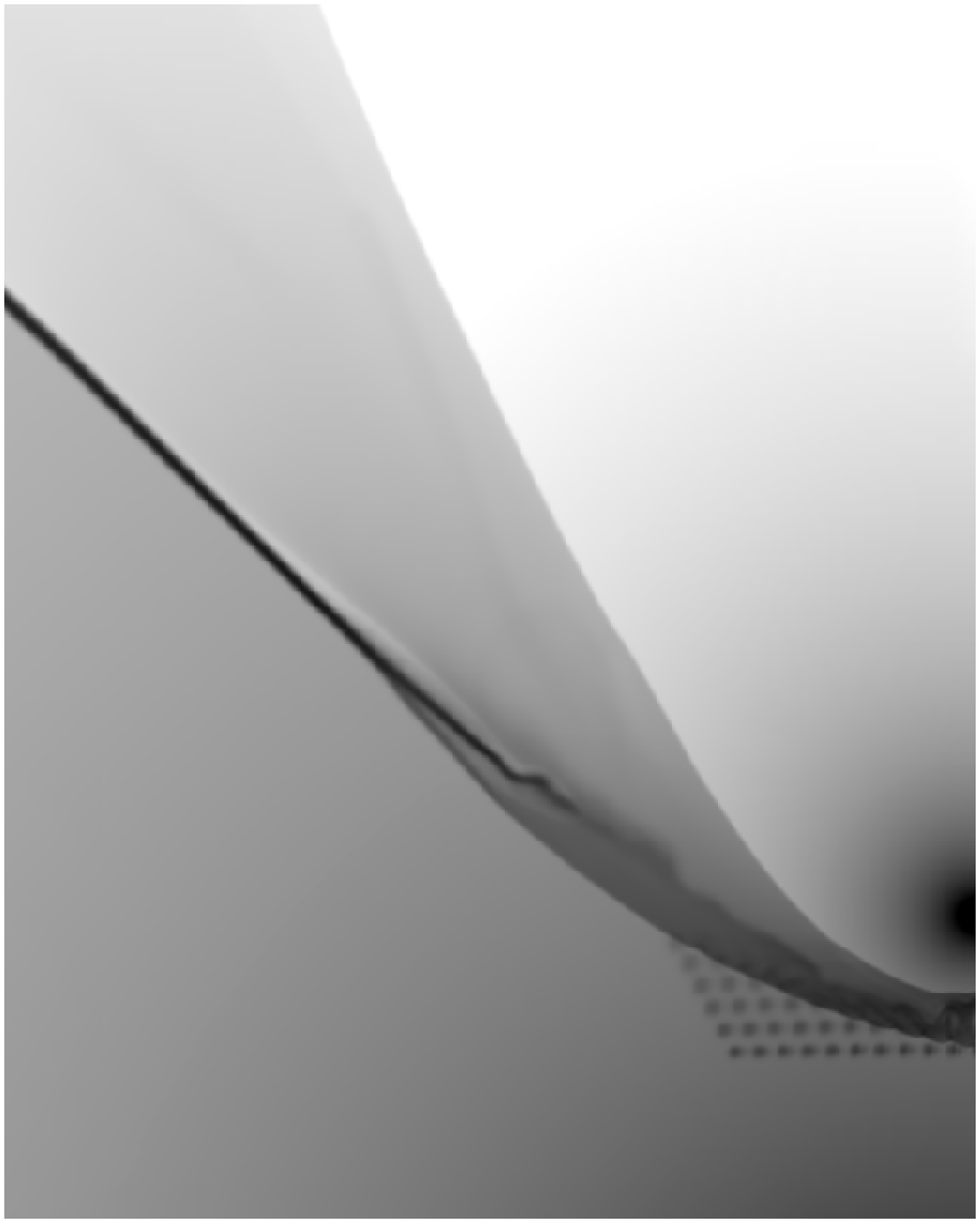}{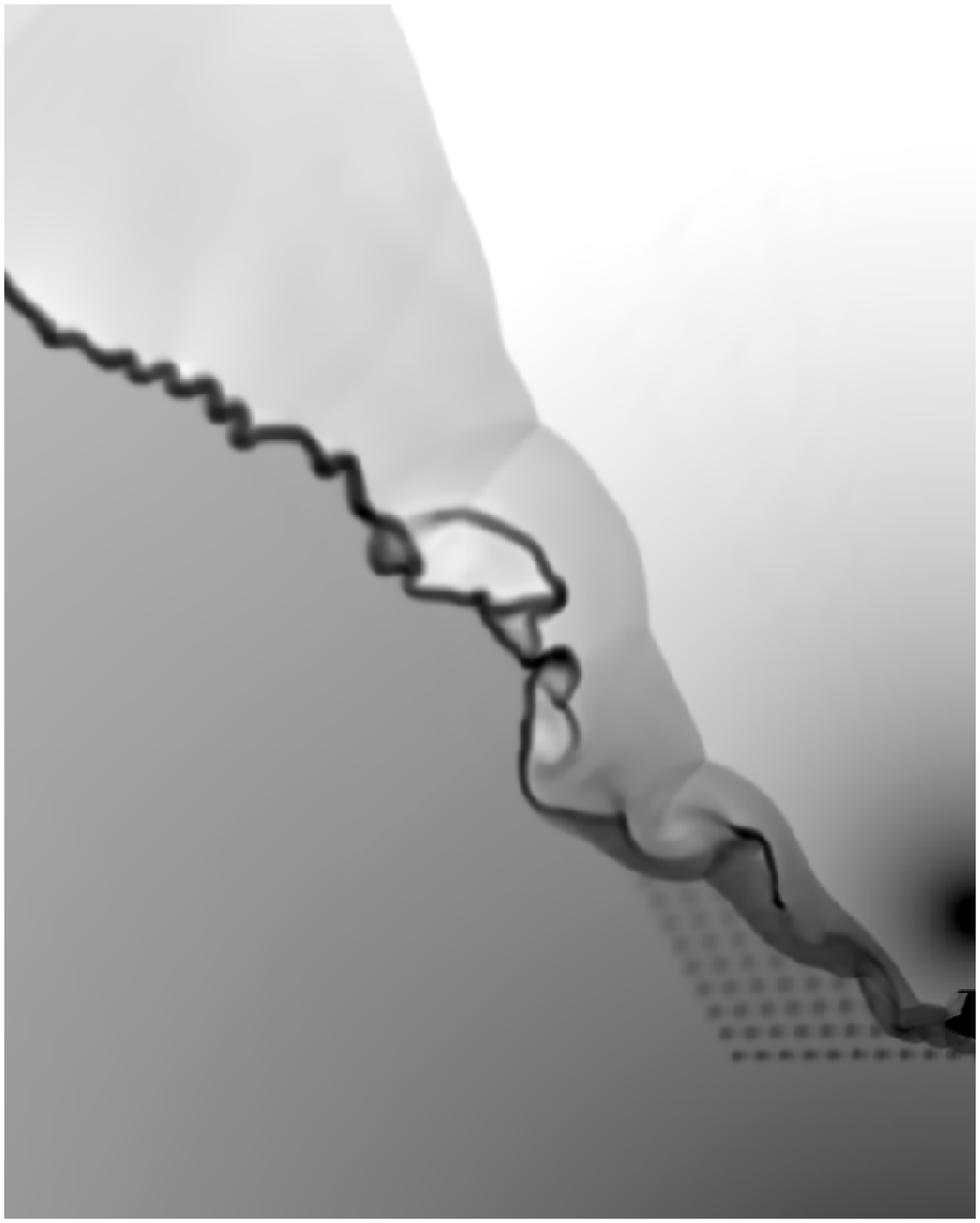}
\caption{Axisymmetric simulation of colliding clumped winds in 
         WR~140 shown in density on a logarithmic scale (white:
         $10^5$~cm$^{-3}$, black: $10^{11}$~cm$^{-3}$ and higher).
         The scale in the horizontal direction is approximately 3~AU.
         Left panel: The first clumps have hit the previously
         stationary interaction zone. Right panel: fully developed
         turbulent interaction zone. Cold, high-density clumps are
         embedded in the hot gas.}
\label{fig:coll_clumped_winds}
\end{figure}

\paragraph{Wide systems} 

We present a 2D simulation of WR140 in the frame of Euler equations
and optically thin radiative cooling. Heat conduction is not
explicitly taken into account. However, the numerical diffusion
inherently present in each numerical simulation mimics also heat
conduction. It is not possible to estimate this effect in detail. It
is, however, likely that around clumps the numerical heat conduction
is stronger than the physically correct one. Figure~2 shows (in
density) the WWIZ of a clumped WR-wind with a smooth
O-star-wind. Again, parameters of the system WR~140 are applied. The
stars are assumed to be in periastron and the orbital movement of the
stars is neglected. The clumps of the WR-wind have a size of
2~R$_{\odot}$ in diameter and are injected in the neighborhood of the
WWIZ. Their density is enhanced by a factor of 3.333... compared to
the density of the stationary wind. The volume occupied by the clumps
is 1/8 of the total volume.

We found that the interaction zone is immediately de-stabilized. Tinny,
high-density knots (hardly visible in the graph) float in the hot
material even in the center of the structure. Mixing does occur in
this simulation. However, the mixing in the simulation is driven by
numerical diffusion which is not controllable. But physical mixing is
likely in the presence of turbulence. Strong compression and mixing
are two ingredients which favor the formation of dust (see e.g.
Cherchneff et al. 2000). For the real formation of dust, temperatures
below typical temperatures of the WWIZ are necessary. Whether this is
possible remains uncertain. However, the reduced strength of the
stellar radiation fields in wide systems, together with occultation
effects described in Walder, Folini \& Motamen (1999) and Folini \&
Walder (this volume) perhaps allow for such cooling. If the system is
too wide, however, even dense clumps may not be able to cool when
shocked. Compression will not take place and thus one ingredient of
dust production is not present.

\paragraph{A note on narrow systems}

Narrow systems have a radiative WWIZ even if smooth flows collide
(Stevens et al. 1992). Such a WWIZ is inherently unstable. In such
systems, the situation is further complicated by the dynamical
importance of the stellar radiation fields (Gayley, Owocki, \& 
Cranmer 1997).

Clumped winds will not generally alter this picture. Minor changes
include probably a more pronounced division between hot and cold
parts: many of the cooled clumps are likely to be denser and some of
the hot gas will have higher temperature and lower density than
predicted by the theory of smooth wind collision. The new generation
of X-ray telescope will allow to test this prediction.

\section{Summary and conclusions}
\label{sec:conclusion}
We have discussed some properties of colliding clumped winds in hot
massive star binaries. We found that in this case the interaction zone
is inherently unstable and turbulent. It is possible, however, that in
a statistical sense, and averaged over typical turbulent time-scales
(not yet known), the interaction zone may again be quasi-stationary.

In wide systems like WR~140, clumped winds result in a partly
radiative interaction zone even in the very center of the system.
This allows the formation of highly condensed knots floating in the
elsewhere hot WWIZ. Such knots can escape evaporation if their size is
on the order of $10^9$~cm. Partial mixing between O-wind and
WR-wind-material is likely due to the turbulence in the dense knots.
The production of dust is favoured by dense, well-mixed
carbon-hydrogen knots.

In narrow systems the overall picture of the WWIZ will not be
altered by clumped wind.

\acknowledgments
The authors benefited from many discussions with the participants of
the workshop. In particular, we are greatful for the comments of Sergej
Marchenko, Yoann Le Teuff, and Andy Pollack.

\end{document}